\documentclass[
aps,
prb,
amsmath,
amssymb,
superscriptaddress,
reprint,
]{revtex4-2}

\usepackage{graphicx}   
\usepackage{dcolumn}    

\usepackage{siunitx}
\DeclareSIUnit\angstrom{\text {Å}}
\usepackage{mathtools}
\usepackage{bbold}
\usepackage{braket}
\usepackage[version=4]{mhchem}
\usepackage{bm}

\usepackage{hyperref}
\usepackage{booktabs}

\usepackage{xcolor}

\begin{document}

\title{Auxiliary field quantum Monte Carlo at the basis set limit: application to lattice constants}

\author{Moritz Humer}
\thanks{Contact author: moritz.humer@univie.ac.at}
\affiliation{Computational Materials Physics \& Vienna Doctoral School in Physics, University of Vienna,  Kolingasse 14-16, A-1090 Vienna, Austria}
\author{Martin Schlipf}
\affiliation{VASP Software GmbH, Berggasse 21/14, A-1090 Vienna, Austria}
\author{Zoran Sukurma}
\affiliation{Institute for Theoretical Physics, TU Wien, Wiedner Hauptstrasse 8-10/E136, A-1040 Vienna, Austria}
\author{Sajad Bazrafshan}
\affiliation{Computational Materials Physics \& Vienna Doctoral School in Physics, University of Vienna, Kolingasse 14-16, A-1090 Vienna, Austria}
\author{Georg Kresse}
\affiliation{VASP Software GmbH, Berggasse 21/14, A-1090 Vienna, Austria}
\affiliation{Computational Materials Physics, University of Vienna, Kolingasse 14-16, A-1090 Vienna, Austria}

\date{\today}

\begin{abstract}
We present a plane-wave (PW) implementation of the auxiliary-field quantum Monte Carlo (AFQMC) method within the projector augmented-wave (PAW) formalism in the Vienna \emph{ab initio} Simulation Package (VASP). By employing an exact inversion of the PAW overlap operator, our approach maintains cubic scaling while naturally operating at the complete basis set limit defined by the PW cutoff. We benchmark this framework by calculating the equilibrium lattice constants and bulk moduli of C, BN, BP, and Si. Our analysis demonstrates that AFQMC systematically corrects the lack of long-range screening in MP2 and the missing higher-order exchange in RPA. We identify RPA as the optimal reference method due to the rapid convergence of the remaining short-range correlations with respect to supercell size. The resulting lattice constants exhibit a mean absolute relative error of \SI{0.14}{\%} relative to experiment, establishing the method as a rigorous benchmark tool for structural properties in condensed matter systems.
\end{abstract}

\maketitle

\section{Introduction}\label{I}
With the success of machine learning~\cite{behler2007,bartok2010,drautz2019,batzner20223,batatia2022mace}, the demand for accurate and reliable reference data has increased dramatically. Since sufficiently high-quality experimental data are not always available, density functional theory~\cite{kohn1965self} (DFT) has become the standard tool for generating such data. With its favorable cubic scaling and access to a wide range of well-established exchange-correlation functionals, DFT offers an unsurpassed balance between efficiency and accuracy. However, due to the approximate nature of available DFT functionals, it is often unclear \emph{a priori} which functional is best suited to a given problem. Indeed, different functionals can yield qualitatively different predictions for structural and thermodynamic properties~\cite{montero2024}. These limitations motivate the development of systematically improvable beyond-DFT approaches for solids.

As a result, low-order perturbational approaches such as M{\o}ller–Plesset perturbation theory~\cite{moller1934note} (MP) and simple diagrammatic many-body methods like the random-phase approximation~\cite{nozieres1958correlation,langreth1977exchange,miyake2002total,fuchs2005describing,furche2008developing} (RPA) have emerged as practical beyond-DFT methods. However, low-order perturbation theories have inherent limitations: MP is inaccurate for strongly polarizable systems with small band gaps, and, more generally, any finite-order perturbation theory diverges for metals at zero temperature~\cite{gruneis2010second,shepherd2013many}.
The RPA, by contrast, accurately describes long-range electronic screening and yields reliable results for a wide range of systems, but neglects exchange-like and ladder diagrams beyond Fock exchange, which limits its accuracy in large-gap materials and in the presence of strong short-range or strong static correlations~\cite{furche2008developing,harl2009accurate,harl2010assessing,schimka2010accurate,lebegue2010cohesive,kaltak2014cubic,garrido2017adsorption,bokdam2017assessing}.

Deterministic wavefunction-based quantum chemistry methods, such as coupled-cluster theory~\cite{vcivzek1966correlation,vcivzek1971correlation}, provide a further increase in accuracy and are actively being developed for periodic systems~\cite{booth2013towards,booth2016plane,liao2016communication,gruneis2017perspective,mcclain2017gaussian,gruber2018applying,irmler2021focal}. However, their steep computational scaling and large memory requirements severely constrain their application. In particular, coupled-cluster singles, doubles, and perturbative triples CCSD(T), often regarded as the ``gold standard'' of quantum chemistry, is a single-reference method and therefore unreliable in the presence of strong static correlations~\cite{bartlett2007coupled}. Moreover, the perturbative triples correction diverges for metallic systems~\cite{shepherd2013many}. Consequently, despite their benchmark accuracy for molecular systems, coupled-cluster methods often remain impractical or unreliable for many applications.

Stochastic approaches offer an appealing alternative, owing to their more favorable scaling and natural parallelizability. Among the quantum Monte Carlo approaches, diffusion Monte Carlo~\cite{anderson1976quantum,ceperley1977monte} (DMC) is routinely applied to solids, but is formally only exact for local pseudopotentials~\cite{foulkes2001quantum,nemec2010benchmark}. The use of non-local pseudopotentials requires additional approximations~\cite{hammond1987valence,mitavs1991nonlocal,casula2006beyond,casula2010size,zen2019new,anderson2021nonlocal}. Furthermore, suitable pseudopotentials are not available for the entire periodic table, in particular for heavier elements or transition metals. DMC has also not yet been formulated within the projector augmented-wave (PAW) framework~\cite{bloechl1994}, which retains near all-electron precision while working with smooth pseudo-orbitals~\cite{humer2022}.

In this work, we focus on the auxiliary-field quantum Monte Carlo (AFQMC) method.
Originally formulated as a path-integral method using the Metropolis algorithm~\cite{hirsch1985two,sugiyama1986auxiliary,white1989numerical,sorella1989novel}, AFQMC was later reformulated into an open-ended random walk in the space of Slater determinants, enabling the introduction of constraints to alleviate the fermionic sign problem~\cite{loh1990sign}. The use of a trial wavefunction to guide the random walk led to the constrained path AFQMC, which proved highly successful for model Hamiltonians~\cite{zhang1995constrained,zhang1997constrained}.
For realistic many-body Hamiltonians, the sign problem generalizes to a phase problem. Zhang and coworkers~\cite{zhang2003quantum} addressed this challenge by developing the phaseless (ph) approximation.
AFQMC has been applied successfully to the Hubbard model~\cite{hubbard1963electron,zheng2017stripe}, and to molecular systems~\cite{al2006auxiliary1,purwanto2008eliminating,purwanto2009excited,purwanto2015auxiliary,al2006auxiliary,al2007study,al2007bond,motta2017towards,landinez2019non,sukurma2023benchmark,sukurma2024toward,shee2019achieving,williams2020direct}. Using Gaussian-type orbitals, ph-AFQMC was compared to DMC in solids~\cite{malone2020systematic}, and applied to NiO~\cite{zhang2018auxiliary}. Suewattana \emph{et al.} formulated ph-AFQMC using plane waves (PWs) together with norm-conserving pseudopotentials~\cite{suewattana2007phaseless}.
PW implementations were later used to study pressure-induced phase transition in silicon~\cite{purwanto2009pressure}, to perform excited state calculations in solids~\cite{ma2013excited}, to investigate the uniform electron gas~\cite{lee2019auxiliary}, and to obtain benchmark charge densities for solids~\cite{chen2021ab}. Recently, the AFQMC framework was also extended to forces and stresses in solids~\cite{chen2023computation}.

In a previous work, we developed a periodic AFQMC implementation importing the Hamiltonian matrix elements from the Vienna \emph{ab initio} Simulation Package~\cite{kresse1996efficient,kresse1999} (VASP) and applied it to prototypical insulators and semiconductors~\cite{taheridehkordi2023phaseless}.
In this work, we directly implement AFQMC within the PAW framework~\cite{bloechl1994} in the PW-based VASP. PWs provide systematic convergence, controlled by a single energy cutoff, and they are independent of atomic positions and chemical species. In practice, pseudopotentials are employed to eliminate the need to explicitly treat tightly bound core electrons. The PAW method has demonstrated high accuracy within DFT for molecules~\cite{paier2005perdew} and solids~\cite{lejaeghere2016reproducibility}, and recently also within correlated wavefunction methods~\cite{humer2022}. Our AFQMC implementation offers a cubic scaling with system size and naturally operates at the basis set limit defined by the plane-wave cutoff, eliminating the need for basis-set extrapolations.

As a first application, we calculate equilibrium lattice constants and bulk moduli for a set of prototypical semiconductors --- C, BN, BP, and Si --- in the thermodynamic limit using MP2, RPA, and AFQMC. We show that MP2 and RPA exhibit systematic errors, while AFQMC-based corrections consistently correct these deficiencies and converge to a common result for sufficiently large supercells.

The remainder of this paper is organized as follows. In Sec.~\ref{Sec.II} we introduce the AFQMC methodology and the PAW formalism. The workflow and computational details are described in Sec.~\ref{Sec.III}, followed by results in Sec.~\ref{Sec.IV}.

\section{Theory}\label{Sec.II}
This section begins with an overview of the AFQMC method formulated in a PW basis, followed by a discussion of the PAW formalism. We emphasize the imaginary-time propagation and the evaluation of the local energy within this framework.

\subsection{AFQMC}\label{Sec.II.B}
The auxiliary-field quantum Monte Carlo (AFQMC) method describes an interacting many-body system by reformulating the time-dependent Schr\"odinger equation in imaginary time. Using the Hubbard--Stratonovich transformation, the interacting problem is mapped onto a non-interacting one coupled to fluctuating auxiliary fields. Unlike DFT, which replaces the many-body interactions with an effective mean-field potential, AFQMC recovers correlation effects through a high-dimensional integral over the auxiliary fields. In practice, this integral is evaluated using Monte Carlo (MC) sampling, implemented as a stochastic random walk in the space of non-orthogonal Slater determinants.

In the following, we derive the second-quantized Hamiltonian used for MC propagation, introduce the imaginary-time projector and the HS transformation, discuss variance-reduction and importance-sampling techniques for controlling the fermionic phase problem, and finally summarize the overall MC algorithm.

\subsubsection{Hamiltonian}\label{Sec.II.B.1}
The derivation of the Hamiltonian follows Ref.~\onlinecite{suewattana2007phaseless}. The Born--Oppenheimer Hamiltonian in a canonical basis and second quantization is
\begin{equation}
    \hat H = \hat{T}_\mathrm{e} + V_\mathrm{ion-ion} + \hat{V}_{\mathrm{e-ion}} + \hat{V}_\mathrm{ee}
    \label{Eq.II.B.1.1}
\end{equation}
with the kinetic energy $\hat{T}_\mathrm{e}$, the ion--ion $V_\mathrm{ion-ion}$, electron--ion $\hat{V}_{\mathrm{e-ion}}$ and electron--electron interaction
\begin{equation}
    \hat{V}_\mathrm{ee} = \sum_{\sigma,\sigma^\prime} \frac{1}{2}\int\! \mathrm{d}^3\bm{r}\, \mathrm{d}^3\bm{r}^\prime\, \frac{\hat{\psi}_\sigma^\dagger(\bm{r})\,\hat{\psi}_{\sigma^\prime}^\dagger(\bm{r}^\prime)\, \,\hat{\psi}_{\sigma^\prime}(\bm{r}^\prime)\,\hat{\psi}_\sigma(\bm{r})}{\left|{\bm{r} - \bm{r}^\prime}\right|}~.
    \label{Eq.II.B.1.2}
\end{equation}
Here and throughout the paper, we use atomic units. The summations run over the spin indices $\sigma$ and $\sigma^\prime$, and $\hat{\psi}_\sigma^\dagger(\bm{r})$ and $\hat{\psi}_\sigma(\bm{r})$ denote the fermionic creation and annihilation field operators at position $\bm{r}$, respectively. $\hat{V}_\text{ee}$ is rewritten in terms of density--density interactions by commuting $\hat{\psi}_\sigma(\bm{r})$ twice to the left using the fermionic anti-commutation relations
\begin{equation}
    \begin{split}
        \centering
        \hat{V}_\mathrm{ee} =& \underbrace{-\sum_{\sigma} \frac{1}{2}\int\! \mathrm{d}^3\bm{r}\, \mathrm{d}^3\bm{r}^\prime\, \hat{\psi}_\sigma^\dagger(\bm{r})\, \frac{1}{\left|{\bm{r} - \bm{r}^\prime}\right|}\, \hat{\psi}_{\sigma}(\bm{r}^\prime)\, \delta(\bm{r} - \bm{r}^\prime)}_{\hat{H}_\mathrm{self}} \\ &+ \underbrace{\frac{1}{2}\int\! \mathrm{d}^3\bm{r}\, \mathrm{d}^3\bm{r}^\prime\, \hat{n}(\bm{r})\, \frac{1}{\left|{\bm{r} - \bm{r}^\prime}\right|}\,  \hat{n}(\bm{r}^\prime)}_{\hat{H}^\prime_{2}},
    \end{split}
    \label{Eq.II.B.1.3}
\end{equation}
where the density operator $\hat{n}(\bm{r})$ is defined as
\begin{equation}
    \centering
    \hat{n}(\bm{r}) = \sum_\sigma \hat{\psi}_\sigma^\dagger(\bm{r})\, \hat{\psi}_\sigma(\bm{r}).
    \label{Eq.II.B.1.4}
\end{equation}
The reformulated two-body Hamiltonian is no longer normal ordered, which manifests in the additional term $\hat{H}_\mathrm{self}$ correcting the self-interactions present in $\hat{H}_2^\prime$. $\hat{H}_\mathrm{self}$ is a one-body operator that is usually treated on the same level as the electron--ion potential. In a canonical orbital basis, $\hat{H}_\mathrm{self}$ involves computations similar to the exact Fock exchange; however, the band summations extend over all orbitals, including virtual (unoccupied) ones. Sec.~\ref{Sec.II.B.4} discusses how we treat this operator in the PAW formalism.

In PW implementations, it is convenient to Fourier transform the density operator to reciprocal space using
\begin{align}
    \centering
    \hat{n}(\bm{G}) =& \int_\Omega \mathrm{d}^3\bm{r}\, \hat{n}(\bm{r})\, e^{-i\bm{G} \cdot \bm{r}} \label{Eq.II.B.1.5} \\
    \hat{n}(\bm{r}) =& \frac{1}{\Omega} \sum_{\bm{G}} \hat{n}(\bm{G})\, e^{i\bm{G} \cdot \bm{r}}, \label{Eq.II.B.1.6}
\end{align}
where $\Omega$ is the volume of the system, and $\bm{G}$ denotes a reciprocal lattice vector. Inserting this into $\hat{H}_2^\prime$ in Eq.~\eqref{Eq.II.B.1.3}, one can write the two-body Hamiltonian in reciprocal space as

\begin{equation}
    \centering
    \hat{H}^\prime_2 = \frac{1}{2} \sum_{\bm{G}\neq 0} \frac{4\pi}{\Omega \bm{G}^2}\, \hat{n}^\dagger(\bm{G})\, \hat{n}(\bm{G}).
    \label{Eq.II.B.1.7}
\end{equation}
Some care needs to be taken with the divergent $\bm{G} = 0$ component, which defines the zero-point of the total energy. It can be shown that the $\bm{G} = 0$ components of the ion--ion, electron--ion, and electron--electron electrostatic interaction cancel each other~\cite{Ihm_1979}.

To improve the efficiency of the MC sampling (see Sec.~\ref{Sec.II.B.3}), it is convenient to reduce the magnitude of the two-body part of the Hamiltonian by a so-called background subtraction~\cite{baer1998shifted,suewattana2007phaseless}. To this end, we write $\hat{H}^\prime_2$ as
\begin{equation}
    \begin{split}
        \centering
        \hat{H}^\prime_2 =& \frac{1}{2} \sum_{\bm{G} \neq 0} \frac{4\pi}{\Omega \bm{G}^2} \left|\hat{n}(\bm{G}) - \bar{n}(\bm{G})\right|^2 \\ &+  \sum_{\bm{G}\neq0} \underbrace{\frac{4\pi}{\Omega \bm{G}^2} \bar{n}(\bm{G})}_{V_\mathrm{H}(\bm{G})} \hat{n}^\dagger(\bm{G}) - \underbrace{\frac{1}{2} \sum_{\bm{G}\neq0} \frac{4\pi}{\Omega \bm{G}^2} \mid \bar{n}(\bm{G}) \mid^2}_{E_\mathrm{H}}, \\
    \end{split}
    \label{Eq.II.B.1.8}
\end{equation}
where the operators $\hat{n}(\bm{G})$ are shifted by their mean-field expectation values
\begin{equation}
    \bar{n}(\bm{G}) = \frac{\bra{\Psi_\mathrm{T}}\hat{n}(\bm{G})\ket{\Psi_\mathrm{T}}}{\braket{\Psi_\mathrm{T} | \Psi_\mathrm{T}}},
    \label{Eq.II.B.1.9}
\end{equation}
evaluated with respect to a trial wavefunction $\ket{\Psi_\mathrm{T}}$ (see Sec.~\ref{Sec.II.B.2} for more details). The additional compensating terms are a one-body term --- the Hartree potential --- and a constant shift --- the Hartree energy $E_\mathrm{H}$.

Finally, the Hamiltonian is sorted according to its one-, two-, and zero-body contributions
\begin{equation}
    \begin{split}
        \centering
        \hat{H}_0 =& V_\mathrm{ion-ion} - E_\mathrm{H}. \\
        \hat{H}_1 =& \hat{T}_\mathrm{e} + \int \mathrm{d}^3\bm{r}~ \left[ V_\mathrm{e-ion}(\bm{r}) + V_\mathrm{H}(\bm{r}) \right]\hat{n}(\bm{r}) + \hat{H}_\mathrm{self} \\
        \hat{H}_2 =& \frac{1}{2} \sum_{\bm{G} \neq 0} \frac{4\pi}{\Omega \bm{G}^2} \left|\hat{n}(\bm{G}) - \bar{n}(\bm{G})\right|^2.
    \end{split}
    \label{Eq.II.B.1.10}
\end{equation}
We note that the first term, $\hat H_0$, is just a constant energy, which naturally commutes with the one-body and two-body terms $\hat{H}_1$ and $\hat{H}_2$, respectively.

\subsubsection{Ground-state projection and Hubbard--Stratonovich transformation}\label{Sec.II.B.2}

AFQMC seeks the ground state $\ket{\Phi_0}$ of an interacting many-body system by solving the imaginary-time Schr\"odinger equation via imaginary-time projection
\begin{equation}
    \centering
    \ket{\Phi_0} = \lim_{n\to\infty} \left[ e^{-\tau(\hat{H}-E_0)}\right]^n\ket{\Psi_\textrm{I}},
    \label{Eq.II.B.2.1}
\end{equation}
where the initial wavefunction $\ket{\Psi_\mathrm{I}}$ must not be orthogonal to the ground state --- \emph{i.e.}, $\braket{\Phi_0 \mid \Psi_\mathrm{I}} \neq 0$. The exponential is split into short-time propagators with an imaginary time step $\tau$. $E_0$ denotes the true ground-state energy; generally unknown, it is usually approximated using an underlying mean-field calculation and refined during the simulation. To isolate the two-body contribution defined by the Hamiltonian $\hat{H}_2$ in Eq.~\eqref{Eq.II.B.1.10}, the short-time propagator is further factorized using a second-order Trotter--Suzuki decomposition~\cite{trotter_1959,suzuki_1976},
\begin{equation}
    \begin{split}
        \centering
        e^{-\tau(\hat{H}-E_0)} = e^{-\frac{\tau}{2}\hat{H}_1} e^{-\tau\hat{H}_2} e^{-\frac{\tau}{2}\hat{H}_1} e^{-\tau(\hat{H}_0-E_0)} + \mathcal{O}(\tau^3),
    \end{split}
    \label{Eq.II.B.2.2}
\end{equation}
introducing an error of order $\mathcal{O}(\tau^3)$. For a detailed discussion on the time-step errors in AFQMC, we refer the interested reader to Ref.~\onlinecite{sukurma2024toward}.

Next, one applies Hubbard--Stratonovich (HS) transformations~\cite{Hubbard_1959,stratonovich_1957} to linearize the two-body propagator
\begin{equation}
    \begin{split}
        \centering
        e^{-\tau\hat{H}_2}
         = \prod_{\bm{G}\neq0} \int \mathrm{d}^2\sigma_{\bm{G}}~ g(\sigma_{\bm{G}}) e^{ -\tau \sigma^*_{\bm{G}} \left[\hat{n}(\bm{G}) - \bar{n}(\bm{G}) \right]} + \mathcal{O}(\tau^2),
    \end{split}
    \label{Eq.II.B.2.3}
\end{equation}
where $g(\sigma_{\bm G})$ denotes a complex normal distribution and $\mathrm{d}^2\sigma_{\bm{G}} = \mathrm{d}x_{\bm{G}} \mathrm{d}\mathrm{y}_{\bm{G}}$ indicates integration over the real and imaginary parts. Because the $\hat{n}(\bm{G})$ are not independent for any pair of reciprocal lattice vectors $\{\bm{G}, -\bm{G}\}$, the complex random field $\sigma_{\bm G}$ must satisfy $\sigma^*_{\bm{G}} = -\sigma_{-\bm{G}}$. This implies that the corresponding real-space field is purely imaginary. To enforce this constraint, we define the $\sigma_{\bm{G}}$ as the Fourier transform of real normal-distributed random variables $\sigma_{\bm{r}} \sim \mathcal{N}(0,1)$
\begin{equation}
    \centering
    \sigma_{\bm{G}} = i\sqrt{\frac{4\pi}{\tau\Omega\bm{G}^2}} \sum_{\bm{r}} e^{-i\bm{G} \cdot \bm{r}} \sigma_{\bm{r}},
    \label{Eq.II.B.2.4}
\end{equation}
where we explicitly include the factor $i$ and choose the normalization to produce the desired variance. The transformation defined in Eq.~\eqref{Eq.II.B.2.3} is exact for pairwise commuting operators; however, if the operators do not commute, the leading error is $\mathcal{O}(\tau^2)$. Physically, the HS transformation replaces a system interacting via a two-body potential with a non-interacting system coupled to the random field $\bm{\sigma}$. Multi-dimensional integration over all fields recovers the full two-body interaction of the interacting system.

To improve the efficiency of the algorithm, Zhang and Krakauer~\cite{zhang2003quantum} introduced an importance sampling transformation
\begin{equation}
    \begin{split}
        \centering
        \int \mathrm{d}\bm{\sigma}~ g(\bm{\sigma}) I(\bm{\sigma},\bar{\bm{\sigma}},\Psi) P(\bm{\sigma} - \bar{\bm{\sigma}}),
    \end{split}
    \label{Eq.II.B.2.5}
\end{equation}
where the stochastic short-time propagator $P(\bm{\sigma})$ is obtained by combining Eqs.~\eqref{Eq.II.B.2.2},~\eqref{Eq.II.B.2.3}, and~\eqref{Eq.II.B.2.4}
\begin{align}
    \centering
    P(\bm{\sigma}) &= e^{-\tau(\hat{H}_0 - E_0)}e^{-\frac{\tau}{2}\hat{H}_1} e^{-\tau \hat{H}_\mathrm{HS}(\bm{\sigma}) } e^{-\frac{\tau}{2}\hat{H}_1} \label{Eq.II.B.2.6} \\
    \hat{H}_\mathrm{HS}(\bm{\sigma}) &= \sum_{\bm{G}\neq0}\sigma^*_{\bm{G}} \left[\hat{n}(\bm{G}) - \bar{n}(\bm{G})\right]. \label{Eq.II.B.2.HS_Hamiltonian}  
\end{align}
For convenience, we introduced the HS-transformed effective one-body Hamiltonian $\hat{H}_\mathrm{HS}(\bm{\sigma})$ and $\bm{\sigma}$ denotes the collection of all $\sigma_{\bm{G}}$. The importance function $I(\bm{\sigma},\bar{\bm{\sigma}},\Psi)$ is given by
\begin{equation}
    \begin{split}
        \centering
        I(\bm{\sigma},\bar{\bm{\sigma}},\Psi) =& \frac{\braket{\Psi_\mathrm{T}| P(\bm{\sigma} - \bar{\bm{\sigma}})|\Psi}}{\braket{\Psi_\mathrm{T}|\Psi}} \\ &\times e^{-\tau\sum_{\bm{G}\neq0}\left(\frac{4\pi}{\Omega\bm{G}^2}\right)^{-1}\left(\bar\sigma^*_{\bm{G}}\sigma_{\bm{G}} - \frac{|\bar\sigma_{\bm{G}}|^2}{2} \right)},
    \end{split}
    \label{Eq.II.B.2.7}
\end{equation}
where $\ket{\Psi}$ is a single Slater determinant and $\ket{\Psi_\mathrm{T}}$ denotes a trial wavefunction  ---  a best guess of the exact many-body wavefunction used to guide the random walk. The trial wavefunction is not restricted to be a single Slater determinant; however, it is often chosen to be identical to the initial wavefunction $\ket{\Psi_\mathrm{I}}$. Finally, note that the integrand in Eq.~\eqref{Eq.II.B.2.5} is shifted by $\bm{\bar \sigma}$, the so-called force bias~\cite{zhang2003quantum}. The HS transformation is invariant under such a shift, and it is chosen such that the variance in the importance function  $I(\bm{\sigma},\bar{\bm{\sigma}},\Psi)$ is minimized
\begin{equation}
    \bar\sigma_{\bm{G}} = -\frac{4\pi}{\Omega\bm{G}^2} \frac{\braket{\Psi_\mathrm{T}|\hat{n}(\bm{G}) - \bar{n}(\bm{G})|\Psi}}{\braket{\Psi_\mathrm{T}|\Psi}}.
    \label{Eq.II.B.2.8}
\end{equation}

\subsubsection{Monte Carlo sampling}\label{Sec.II.B.3}
In practice, the high-dimensional integral in Eq.~\eqref{Eq.II.B.2.3} is evaluated through an MC procedure: introducing a walker ensemble, the integral is sampled by a single configuration of the auxiliary field $\bm{\sigma}$ for each walker $w$ and time step $k$. Each walker $\{\ket{\Psi_k^w},W_k^w e^{i\theta_k^w}\}$ in the ensemble is described by a single-determinant wavefunction $\ket{\Psi^w_k}$ and a weight $W^w_k e^{i\theta^w_k}$, which is generally complex. Since the propagator $P(\bm{\sigma} - \bar{\bm{\sigma}})$ defined in Eq.~\eqref{Eq.II.B.2.6} is a one-body operator, Thouless' theorem ensures that the walkers remain single Slater determinants upon propagation~\cite{thouless1960stability,thouless1961vibrational}.
At step $k=0$, each walker in the ensemble is initialized as
\begin{equation}
    \centering
    \ket{\Psi_0^w} = \ket{\Psi_\mathrm{I}};\hspace{1cm} W_0^we^{i\theta_0^w} = \braket{\Psi_\mathrm{T}|\Psi_\mathrm{I}},
    \label{Eq.II.B.3.1}
\end{equation}
where $\ket{\Psi_\mathrm{I}}$ is an initial single Slater determinant  ---  usually an underlying mean-field wavefunction --- and $\ket{\Psi_\mathrm{T}}$ denotes the trial wavefunction discussed in Sec.~\ref{Sec.II.B.2}. Then, for each time step $k$ along the projection defined in Eq.~\eqref{Eq.II.B.2.1}, the many-body wavefunction $\ket{\Phi_k}$ is represented by
\begin{equation}
    \centering
    \ket{\Phi_k} = \sum_w W_k^w e^{i\theta_k^w} \frac{\ket{\Psi_k^w}}{\braket{\Psi_\mathrm{T}|\Psi_k^w}}.
    \label{Eq.II.B.3.2}
\end{equation}

Along the trajectory, the walkers and their associated weights are updated within the phaseless approximation~\cite{zhang2003quantum} proposed by Zhang and Krakauer
\begin{align}
    \centering
    \ket{\Psi_{k+1}^w} &= P(\bm{\sigma} - \bar{\bm{\sigma}})\ket{\Psi_{k}^w} \label{Eq.II.B.3.3} \\
    W_{k+1}^w &= \left| I(\bm{\sigma},\bm{\bar\sigma},\Psi_k^w) \right|  W_{k}^w \max \left (0,\cos(\Delta\theta_k^w) \right) \label{Eq.II.B.3.4} \\
    \theta_{k+1}^w &= 0, \label{Eq.II.B.3.5}
\end{align}
where the phase change $\Delta\theta_k^w$ is
\begin{equation}
    \Delta\theta_k^w = \Im \log \frac{\braket{\Psi_\mathrm{T}|\Psi_{k+1}^w}}{\braket{\Psi_\mathrm{T}|\Psi_k^w}}.
    \label{Eq.II.B.3.6}
\end{equation}
Designed to mitigate the fermionic phase problem, the phaseless approximation imposes a universal gauge on the phase of the walker weights via the trial wavefunction $\ket{\Psi_\mathrm{T}}$, resulting in real-valued weights. This keeps the algorithm stable; however, it introduces a systematic error, known as the phaseless error. It is understood that improving the trial wavefunction or employing release constraint techniques systematically reduces the phaseless error~\cite{Mahajan_Selected,Xiao_Interfacing}.

Finally, after an equilibration phase, the ground-state energy is estimated by the time-averaged estimator
\begin{equation}
    \centering
    \bar{E} = \frac{\sum_{kw} W_k^w E_\mathrm{loc}(\Psi_k^w)}{\sum_{kw} W_k^w},
    \label{Eq.II.B.5.1}
\end{equation}
where the local energy $E_\mathrm{loc}$ is given by the mixed estimator
\begin{equation}
    \begin{split}
        \centering
        E_\mathrm{loc}(\Psi_k^w) =& \frac{\braket{\Psi_\mathrm{T}|\hat{H}|\Psi_k^w}}{\braket{\Psi_\mathrm{T}|\Psi_k^w}}.
    \end{split}
    \label{Eq.II.B.5.2}
\end{equation}
Note that the mixed estimator is only exact for observables that commute with the propagator and hence with the Hamiltonian. The detailed local energy evaluation within the PAW formalism is discussed in Sec.~\ref{Sec.II.B.5}.

\subsection{The PAW formalism}\label{Sec.II.A}

Here and in the following, all operators are one-body operators and, if enclosed between orbitals $\tilde{\psi}_n$, they act in a (one-body) Hilbert space. In the projector augmented-wave (PAW) formalism~\cite{bloechl1994}, the variational quantities are the so-called pseudo orbitals $\tilde{\psi}_n$, expanded in a PW basis
\begin{align}
    \centering
        \braket{\bm{r} \mid \tilde{\psi}_n} &=~ \sum_{\bm{G}} \mathbf{C}_{\bm{G}n} \braket{\bm{r} \mid \bm{G}} \label{Eq.II.A.1} \\
        \braket{\bm{r} \mid \bm{G}} &=~\frac{1}{\sqrt{\Omega}}e^{i\bm{G}\cdot\bm{r}}, \label{Eq.II.A.planewave}
\end{align}
where $\Omega$ is the volume of the Wigner--Seitz cell and $\ket{\bm{G}}$ denotes a reciprocal lattice vector. The pseudo orbitals $\tilde{\psi}_n$ are related to the all-electron (AE) orbitals $\psi_n$ via the linear transformation
\begin{equation}
    \begin{split}
        \centering
        \ket{\psi_n} &= \hat{T} \ket{\tilde{\psi}_n}  \\
        &= \ket{\tilde{\psi}_n} + \sum_i \left(\ket{\phi_i} - \ket{\tilde{\phi}_i} \right) \braket{\tilde{p}_i | \tilde{\psi}_n},
    \end{split}
    \label{Eq.II.A.2}
\end{equation}
where $\tilde{p}_i$, $\phi_i$, and $\tilde\phi_i$ are projectors, AE and pseudo partial waves, respectively~\cite{bloechl1994}. The index $i$ is a shorthand for the atomic site $\bm{R}$, the angular momentum quantum numbers $L = (l,m)$, and an additional index $k$ for the reference energy $\epsilon_{kl}$. The $\phi_i$ are the solutions of the radial Schr\"odinger equation for a non-spin-polarized reference atom at energies $\epsilon_{kl}$. Outside the atomic spheres, the $\tilde\phi_i$ coincide with the $\phi_i$. Within the atomic spheres, the $\tilde\phi_i$ are chosen to be computationally convenient (smooth) functions. The projector functions $\tilde p_i$ are localized in the atomic sphere and should satisfy $\sum_i \ket{\tilde\phi_i}\bra{\tilde p_i}=\mathbb{1}$ within each atomic sphere. This ensures that the one-center expansion of the pseudo orbital $\sum_i \ket{\tilde\phi_i}\braket{\tilde p_i | \tilde\psi_n}$ is identical to the pseudo orbital $\tilde\psi_n$ within the atomic sphere. The one-center expansion $\sum_i \ket{\phi_i}\braket{\tilde p_i | \tilde\psi_n}$ restores the exact shape of the AE orbital within the atomic sphere on the radial support grid.

Unlike the orthogonal AE orbitals, the pseudo orbitals are $S$-orthogonal
\begin{equation}
    \centering
    \braket{\tilde{\psi}_n \mid \hat{S} \mid \tilde{\psi}_m} = \delta_{nm},
    \label{Eq.II.A.7}
\end{equation}
where  $\hat S = \hat{T}^\dagger \hat{T}$ denotes the PAW overlap operator. Using Eq.~\eqref{Eq.II.A.2} together with the completeness relation $\sum_i \ket{\tilde{\phi_i}}\bra{\tilde{p}_i} = \mathbb{1}$, valid within each atomic sphere, it can be written as
\begin{equation}
    \centering
    \begin{split}
        \hat{S} &= \mathbb{1} + \sum_{i,j} \ket{\tilde{p}_i} \left(\braket{\phi_i | \phi_j} - \braket{\tilde{\phi}_i | \tilde{\phi}_j}\right) \bra{\tilde{p}_j} \\
                &= \mathbb{1} + \sum_{i,j} \ket{\tilde{p}_i} \mathbf{Q}_{ij} \bra{\tilde{p}_j}.
    \end{split}
    \label{Eq.II.A.8}
\end{equation}
The augmentation charges $\mathbf{Q}_{ij}$ are block diagonal in the atomic site index $\bm{R}$ because the projectors are localized within the sphere surrounding each atomic site. Finally, the pseudo orbitals are the solutions to a generalized one-body eigenvalue problem
\begin{equation}
    \centering
    \hat{H}\ket{\tilde{\psi}_n} = \epsilon_n \hat{S}\ket{\tilde{\psi}_n}.
    \label{Eq.II.A.9}
\end{equation}

\subsubsection{Energy measurement}\label{Sec.II.B.5}

The local energy, defined in Eq.~\eqref{Eq.II.B.5.2}, is evaluated using the generalized Wick's theorem~\cite{Balian_Nonunitary} and split into a one-body contribution, a Hartree-like, and an exchange-like term. The one-body contribution to the local energy for a specific walker with the determinant $\Psi_k^w$ is
\begin{equation}
    \centering
    E_1(\Psi_k^w) = \frac{\braket{\Psi_\mathrm{T}|\hat{T}_\mathrm{e} + \hat V_\mathrm{e-ion} + V_\mathrm{ion-ion}|\Psi_k^w}}{\braket{\Psi_\mathrm{T}|\Psi_k^w}}~.
    \label{Eq.II.B.5.3}
\end{equation}
The denominator implies that the walker orbitals need
to be aligned with the orbitals in the trial wavefunction for any energy evaluation. To achieve this, one calculates the overlap between the two sets of wavefunctions and applies its inverse to the walker orbitals.
If the determinants $\Psi_k^w$ and $\Psi_\mathrm{T}$ consist of pseudo orbitals $\{ \tilde\psi_{m} \vert m=1,..,N_{\mathrm{occ}} \}$ and
$\{ \tilde\psi^T_{m} \vert n=1,..,N_{\mathrm{occ}} \}$, respectively,  their overlap is evaluated as
\begin{equation}
     \braket{\tilde{\psi}^T_n \mid\hat{S} \mid \tilde{\psi}_m} = \mathbf{O}_{nm}
\end{equation}
In subsequent energy evaluations, the orbitals 
$\mathbf{O}^{-1}  \tilde{\psi}$ need to be used. To avoid an overly cumbersome notation, we will continue to use the symbol $\tilde{\psi}$ for these aligned orbitals. The one-body term is straightforward to evaluate and not discussed further.
The electron--electron interaction is given by a direct Hartree-like term
\begin{equation}
    \centering
    \begin{split}
        E_\mathrm{d}(\Psi_k^w) =& \frac{1}{2} \sum_{\bm{G}} \frac{4\pi}{ \Omega \bm{G}^2} \sum_m^\mathrm{occ} \bra{\tilde\psi^\mathrm{T}_{m}} \hat n(\bm{r}) e^{i\bm{G}\cdot\bm{r}} \ket{\tilde\psi_{m}} \\ &\times \sum_n^\mathrm{occ} \bra{\tilde\psi^\mathrm{T}_{n}} \hat n(\bm{r}^\prime) e^{-i\bm{G}\cdot\bm{r}^\prime} \ket{\tilde\psi_{n}}
    \end{split}
    \label{Eq.II.B.5.4}
\end{equation}
and an exchange-like term
\begin{equation}
    \centering
    \begin{split}
        E_\mathrm{ex}(\Psi_k^w) =& -\frac{1}{2} \sum_{\bm{G}} \frac{4\pi}{\Omega \bm{G}^2}  \sum_{m,n}^\mathrm{occ} \bra{\tilde\psi^\mathrm{T}_{m}} \hat n(\bm{r}) e^{i\bm{G}\cdot\bm{r}} \ket{\tilde\psi_{n}} \\ &\times \bra{\tilde\psi^\mathrm{T}_{n}} \hat n(\bm{r}^\prime) e^{-i\bm{G}\cdot\bm{r}^\prime} \ket{\tilde\psi_{m}},
    \end{split}
    \label{Eq.II.B.5.5}
\end{equation}
where spin degrees of freedom are neglected. Here $\hat n(\bm{r})$ is the one-body density operator in the PAW method. The band summations $m$ and $n$ are performed only over occupied orbitals. The k-point summations are omitted, since the presented implementation supports only a single k-point and is thus restricted to zero momentum transfer ($\bm{q} = 0$). The spatial integrals in Eqs.~\eqref{Eq.II.B.5.4} and \eqref{Eq.II.B.5.5} are restricted to the primitive unit cell and are defined as follows:
\begin{widetext}
\begin{equation}
    \bra{\tilde\psi^\mathrm{T}_{n}}\hat n(\bm{r}) e^{-i\bm{G}\cdot\bm{r}} \ket{\tilde\psi_{m}} = \int_{\Omega_0} \mathrm{d}^3\bm{r}~ e^{-i\bm{G}\cdot\bm{r}} 
    \Bigg( \tilde\psi^\mathrm{T}_{n}(\bm{r})  \tilde\psi_{m}(\bm{r}) +
   \sum_{i,j} \braket{\tilde\psi_{n} | \tilde{p}_i} \left(\braket{\phi_i |\bm{r}| \phi_j} - \braket{\tilde{\phi}_i | \bm{r}|  \tilde{\phi}_j}\right) \braket{\tilde{p}_j | \tilde\psi_{m}} \Bigg).
\end{equation}
\end{widetext}
The one-body contribution is evaluated fully consistently with the PAW formalism, but as common to the PAW method the exact augmentation charge density $\braket{\phi_i |\bm{r}| \phi_j} - \braket{\tilde{\phi}_i | \bm{r}|  \tilde{\phi}_j}$ is approximated by a pseudo charge density $\mathbf{Q}_{ij}(\bm{r})$~\cite{shishkin2006implementation}. Specifically, the direct and exchange contributions to the local energy are evaluated only on the PW grid; the PAW one-center terms are not implemented. This strategy is also adopted in other correlated methods in VASP (\emph{e.g.}, RPA and GW). The potential errors are corrected by restoring the moments as well as the exact shape of the all-electron density on the PW grid~\cite{shishkin2006implementation,unzog2022x}.
Furthermore, one can show that energy changes related to changes of the one-body density matrix from the mean field reference to the AFQMC calculation, are exactly described to linear order by the one-body term Eq.~\eqref{Eq.II.B.5.3}. 
For the materials investigated in this work, we expect the inclusion of the one-center terms to have only a minor impact on the equilibrium lattice constant. To confirm this, we compared MP2 correlation energies obtained with (i) exact one-center terms, (ii) shape restoration, and (iii) using only the PW grid contribution. The resulting equilibrium lattice constants differed by less than \SI{0.1}{\%}.

\subsubsection{Imaginary-time propagation in the PAW method}\label{Sec.II.B.4}

Related to Eq.~\eqref{Eq.II.A.9}, in the presence of an overlap operator, the imaginary-time Schr\"odinger equation becomes
\begin{align}
    &-\hat{S} \frac{\mathrm{d}}{\mathrm{d}\tau} \ket{\tilde{\psi}_n(\tau)}  = \hat{H}\ket{\tilde{\psi}_n(\tau)} \Leftrightarrow
       \label{Eq.II.A.time1}
    \\
     &-\frac{\mathrm{d}}{\mathrm{d}\tau} \ket{\tilde{\psi}_n(\tau)} = \hat{S}^{-1} \hat{H}\ket{\tilde{\psi}_n(\tau)}.
     \label{Eq.II.A.time2}
\end{align}
A convenient way to deal with the overlap operator is to express the pseudo orbitals in a time-independent basis $\ket{\tilde{\psi}_k}$ of some mean field reference Hamiltonian, either the Hartree--Fock (HF) or the DFT Hamiltonian
\begin{equation}
     \ket{\tilde{\psi}_n(\tau)} = \sum_m  \mathbf{A}_{nm}(\tau)\ket{\tilde{\psi}_m}, 
    \label{Eq.II.A.ansatz}
\end{equation}
where the orbital basis must be $S$-orthogonal, as specified in Eq.~\eqref{Eq.II.A.7}. If the Hamiltonian is expressed in an $S$-orthogonal basis 
\begin{equation}
     \mathbf{H}_{nm} = \braket{\tilde{\psi}_n \mid \hat{H} \mid \tilde{\psi}_m}, 
     \label{eq:matrixcanoncical}
\end{equation}
the imaginary-time Schr\"odinger equation can be written as a simple matrix equation, avoiding the inversion of the overlap operator: inserting the ansatz Eq.~\eqref{Eq.II.A.ansatz} into Eq.~\eqref{Eq.II.A.time1} and multiplying with the dual basis states $\bra{\tilde{\psi}}$ yields 
\begin{equation}
    \centering
    -\frac{\mathrm{d}}{\mathrm{d}\tau } \mathbf{A}(\tau) = \mathbf{H} \mathbf{A}(\tau).
    \label{Eq.II.A.timematrix}
\end{equation}

Often, the operators and the matrices --- \emph{i.e.}, the operators evaluated in the canonical basis ---
are interchangeable. However, in the PAW method, the imaginary-time evolution operator in the operator form is given by [compare Eq.~\eqref{Eq.II.A.time2}]
\begin{equation}
    e^{-\tau\hat{S}^{-1} \hat{H}}.\label{Eq.II.B.2_time}
\end{equation}
Thus, in principle, in any emerging propagator in Sec.~\ref{Sec.II.B}, the Hamiltonian in the exponent needs to be replaced by $\hat{S}^{-1} \hat{H}$. This, however, is hard to work with, since this expression involves the expensive-to-evaluate inversion of an operator. A solution to this issue is discussed later in this section alongside the propagation with $\hat{H}_\mathrm{HS}$ [compare the splitting of the short-time propagator defined in Eq.~\eqref{Eq.II.B.2.6}].

Alternatively, one can work in the canonical orbital basis propagating the coefficient matrices $\mathbf{A}$ using the operator [compare Eq.~\eqref{Eq.II.A.timematrix}]
\begin{equation}
    e^{-\tau\mathbf{H}}.
\end{equation}
This is equally involved for large basis sets. Nevertheless, we adopt this approach when propagating using $\hat{H}_1$.

As is common in PW codes, the orbital coefficients $\mathbf{C}_{\bm{G}n}(\tau)$, which are defined in Eq. \eqref{Eq.II.A.1}, are stored in a matrix denoted by the symbol $\mathbf{C}(\tau)$ for each walker and time step. 
To indicate this dependence, we have explicitly included a functional dependence on  $\tau$ (suppressing the walker index). Note that only occupied orbitals need to be considered in the time propagation. For clarity, the dimensions of the matrices discussed in this section are summarized in Table~\ref{Tab.II.B.4.1}.
\begin{table}
    \caption{\label{Tab.II.B.4.1} List of the dimension of each matrix 
    in the PW code.}
    \begin{ruledtabular}
    \begin{tabular}{lllllllll}
        number of occupied orbitals & $N_\mathrm{occ}$   \\
        number of PW coefficients & $N_\mathrm{pw}$ \\
        matrix $\mathbf{C} (\tau)$ & $N_\mathrm{pw}\times N_\mathrm{occ}$ \\
        matrix $\mathbf{C} (0)$ & $N_\mathrm{pw}\times N_\mathrm{pw}$ \\
        matrix $\mathbf{A} (\tau)$ & $N_\mathrm{pw}\times N_\mathrm{occ}$ \\
    \end{tabular}
    \end{ruledtabular}
\end{table}

We also store the initial wavefunction determinant as a set of coefficients
$\mathbf{C}_{\bm{G}n}(0)$, where 0 indicates the set of orbitals at the initial time 0. We have chosen a complete basis for these orbitals, for example, by diagonalizing a mean-field Hamiltonian (usually the HF or DFT Hamiltonian) exactly.
This means that the matrix $\mathbf{C}(0)$ is a square matrix (see Table~\ref{Tab.II.B.4.1}), which observes 
$S$-orthogonality [compare Eq.~\eqref{Eq.II.A.7}]:
\begin{flalign}    
  & \sum_{\bm{G}\bm{G}'}  \mathbf{C}_{\bm{G}s}(0)^*  \hat{\mathbf{S}}_{\bm{G},{\bm{G}'}}   \mathbf{C}_{\bm{G'}r}(0) = \delta_{sr} \label{Eq.II.B.4.1} \\
   &  \hat{\mathbf{S}}_{\bm{G},{\bm{G}'}} =   \braket{\bm{G} \mid \hat{S} \mid \bm{G}'}.
     \label{Eq.II.B.4.S_matrix}
\end{flalign}  
Here and in the following, indices $s$ and $r$ go over all the basis functions, including unoccupied states.  To distinguish matrix representations in the orbital and the PW basis, we include the operator accent ``hat'' for matrices in the PW basis: $\hat{\mathbf{S}}$ and $\hat{\mathbf{H}}$ are the overlap operator and Hamiltonian evaluated between PWs,
as opposed to matrices $\mathbf{H}$ evaluated in an $S$-orthogonal basis [compare Eq.~\eqref{eq:matrixcanoncical}].

In the present implementation, the one-body Hamiltonian $\hat{H}_1$ is evaluated in an $S$-orthogonal basis, since this enables exact exponentiation without error: the condition number of the matrix $\hat{H}_1$, \emph{i.e.}, the quotient of the largest and smallest eigenvalue, is usually huge. If we want to use large time steps, iterative algorithms are not competitive. Additionally, the self-interaction term $\hat{H}_\mathrm{self}$ is difficult to implement in the PAW formalism, whereas it is straightforward to evaluate in a canonical basis (see Sec.~\ref{Sec.II.B.1} ).
To switch from the PW coefficients $\mathbf{C}(\tau)$ to the coefficient matrix $\mathbf{A}(\tau)$, one needs to calculate
\begin{equation}
    \mathbf{A}_{sn}(\tau) = \braket{\tilde{\psi}_s(0) \mid \hat S  \tilde{\psi}_n(\tau)}. 
\end{equation}
Evaluating $\mathbf{A}_{sn}(\tau)$ is most conveniently done by inserting a complete set of PWs before the operator $\hat S$ yielding
\begin{equation}
   \mathbf{A}_{sn}(\tau) = \sum_{\bm{G}}  \mathbf{C}_{\bm{G}s}(0)^*  \braket{\bm{G} \mid \hat{S} \tilde{\psi}_n(\tau)}.
   \label{Eq.II.B.4.coeff_mat}
\end{equation}
In the VASP code, evaluating the action of the operator $\hat{S} \tilde{\psi}_n(\tau)$ and subsequent projection onto PWs is a standard and fast operation, scaling as the product of the number of occupied orbitals $N_\mathrm{occ}$ and the number of PW coefficients $N_\mathrm{pw}$.
Therefore, the time-consuming step is the matrix multiplication with $\mathbf{C}^\dagger(0)$, which scales cubically with the system size $N_\mathrm{occ} \times N_\mathrm{pw}^2$ when performed for all time-propagated orbitals.
The operator  $\hat{H}_1$ is evaluated at the beginning of the AFQMC using a complete $S$-orthogonal basis
\begin{equation}
    \mathbf{H}_{1,sr} = \braket{\tilde{\psi}_s(0) \mid \hat H_1 \mid \tilde{\psi}_r(0)},
\end{equation}
and the matrix is then diagonalized and exponentiated (by exponentiating the eigenvalues).

After the propagation, the new coefficients $\mathbf{B}(\tau_0+\frac{\tau}{2}) = \exp( -\frac{\tau}{2} \mathbf{H}_1) \mathbf{A}(\tau_0) $ are  transformed back from the $S$-orthogonal basis to the PW basis; the transformation
\begin{equation}
     \mathbf{C}_{\bm{G}k}\left( \tau_0 +\frac{\tau}{2} \right)  \leftarrow   \sum_{s}  \mathbf{C}_{\bm{G}s}(0) \bm{B}_{s k}\left( \tau_0+ \frac{\tau}{2} \right)
\end{equation}
yields the final time-propagated PW coefficients. As indicated above, the previously stored PW coefficients are overwritten by the new ones.
Starting from a set of PW coefficients $\mathbf{C}\left( \tau_0 \right)$, we can compactly write the full propagator as follows:
\begin{equation}
    \mathbf{C}\left(\tau_0 + \frac{\tau}{2}\right) \leftarrow \mathbf{C}(0) \exp\left(-\frac{\tau}{2} \mathbf{H}_1\right) \mathbf{C}^\dagger(0) \hat{\mathbf{S}}  \mathbf{C}\left( \tau_0 \right).
     \label{equ:timepropmatrix}
\end{equation}
In practice, the walker orbitals are never transformed to an $S$-orthogonal basis during propagation. Instead, the square matrix $\mathbf{C}(0) \exp(-\frac{\tau}{2} \mathbf{H}_1) \mathbf{C}^\dagger(0)\hat{\mathbf{S}}$ is calculated once and applied subsequently. We found the total compute time for this step acceptable, given that all operations were realised using highly performant BLAS Level 3 routines. However, some rank-compression algorithms might be advantageous for future applications.

Unlike $\hat{H}_1$, the HS-transformed Hamiltonian $\hat{H}_\mathrm{HS}$ is stochastic, as it is evaluated with a different auxiliary field for each walker and time step. It is thus prohibitively expensive to compute and exponentiate its matrix elements in the $S$-orthogonal basis. Instead, we use an iterative algorithm working in the PW basis: the propagator for a generalized eigenvalue equation is given by  Eq.~\eqref{Eq.II.B.2_time}.
In the PW basis, the operators are simply replaced by their corresponding matrices evaluated in the PW basis:
\begin{equation}
    \centering
    e^{-\tau \hat{\mathbf{S}}^{-1} \hat{\mathbf{H}}_\mathrm{HS}(\bm{\sigma})}.
\end{equation}
In the appendix, we show that propagation in the PW basis is strictly equivalent to propagation in the $S$-orthogonal basis. 
The action of the propagator on the orbitals is calculated with iterative methods such as Taylor expansion or Krylov subspace exponentiation (see Ref.~\onlinecite{sukurma2024toward} for more details). All these approaches rely on repeated application of the operator $\hat{S}^{-1}\hat{H}_\mathrm{HS}(\bm{\sigma})$ to the orbitals. 
In VASP's iterative algorithms for determining the \textit{ground state}, the calculation of the inverse of the overlap matrix is avoided, but instead the orbitals that enter in the iterative Krylov subspace are made $S$-orthogonal by Gram--Schmidt-like procedures. Unfortunately, we found that 
simply keeping the iterative subspace $S$-orthogonal does not suffice to construct an efficient iterative algorithm for the propagator. Instead, it is essential to either calculate the inverse of the $\hat S$ operator exactly, or to use another iterative procedure 
to solve the linear equation $\hat{S} \bm{x} = \bm{a} $ for $\bm{x}$. However, nested iterative procedures are too slow and unwieldy for designing an iterative algorithm for the exponential.

To efficiently evaluate the action of $\hat{S}^{-1}$, we make an ansatz analogous to Eq.~\eqref{Eq.II.A.8}
\begin{equation}
    \centering
    \hat{S}^{-1} = \mathbb{1} + \sum_{i,j} \ket{\tilde{p}_i} \bar{\bm{Q}}_{ij} \bra{\tilde{p}_j},
    \label{Eq.II.B.4.4}
\end{equation}
where the composite index $i$ labels the atomic site $\bm{R}$, the angular momentum quantum numbers $L=(l,m)$, and reference energy index $k$. Imposing the condition $\hat{S}^{-1} \hat{S} = \mathbb{1}$, the coefficients $\bar{\bm{Q}}$ are obtained as
\begin{equation}
    \centering
    \bar{\bm{Q}} = - \bm{Q}\left(\mathbb{1} + \bm{M} \Re[\bm{Q}] \right)^{-1} \quad\textrm{with}\quad \bm{M}_{ij} = \braket{\tilde{p}_i|\tilde{p}_j}.
    \label{Eq.II.B.4.5}
\end{equation}
While the augmentation charge matrix $\bm{Q}$ is block diagonal in the atomic site index, the coefficient matrix $\bar{\bm{Q}}$ is generally not. The use of the real part of $\bm{Q}$ in the inverse ensures that $\hat{S}^{-1}$ is Hermitian. The coefficients $\bar{\bm{Q}}$ are pre-calculated, and the action of $\hat{S}^{-1}$ can then be evaluated with similar routines as the action of $\hat{S}$; however, generalized for non-block diagonal coefficient matrices.
The total cost of this step also increases cubically, but the matrix dimension of the inverse overlap operator is typically a factor \numrange{5}{10} times smaller than the number of PWs. This reduces the cost by a factor \numrange{25}{100} compared to the computational cost of transforming the orbitals from the PW basis to the $S$-orthogonal basis, as used in Eq.~\eqref{equ:timepropmatrix}.

A final note is in place. The present AFQMC implementation naturally includes excitations in all $\bm{G}$ components within the PW cutoff sphere --- corresponding to calculations in the basis set limit defined by the PW cutoff.

\section{Computational details}\label{Sec.III}
All results presented in this work were obtained using VASP~\cite{kresse1999,paier2005perdew} within the PAW framework~\cite{bloechl1994} using a PW basis. The GW-type PAWs were selected because they yield accurate scattering properties well above the vacuum level. The details of the pseudopotentials --- including the valence electrons, the matching core radii, and the cutoff for the local potential --- are listed in Table~\ref{Tab.III.1}.
\begin{table}
\caption{\label{Tab.III.1} List of the PAW potentials used in the present work. Shown are the orbitals that are treated as valence orbitals and the label of the respective PAW potential in the VASP database. The radial cutoffs for each angular momentum quantum number are specified as $n\times r_{\rm cut}$, where $n$ specifies the number of projectors and $r_{\rm cut}$ is the radial cutoff in atomic units for the specific angular momentum quantum number. The local potential corresponds to the all-electron potential, which is replaced by a local pseudopotential below the radius $r_{\rm core}$ (a.u.).}
\begin{ruledtabular}
\begin{tabular}{lllllllll}
atom      & valence   & label & s             &  p            &   d           &   f          &$r_{\rm core}$ \\
\midrule
B         & $2s^22p^1$    & GW  & 2$\times$ 1.5 & 2$\times$ 1.7 & 1$\times$ 1.7 &              &1.7 \\
C         & $2s^22p^2$    & GW  & 2$\times$ 1.2 & 2$\times$ 1.5 & 1$\times$ 1.5 &              &1.5 \\
N         & $2s^22p^3$    & GW  & 2$\times$ 1.3 & 2$\times$ 1.5 & 1$\times$ 1.5 &              &1.5 \\
Si        & $3s^23p^2$    & GW  & 2$\times$ 1.9 & 2$\times$ 1.9 & 2$\times$ 1.9 & 1$\times$1.9 &1.9 \\
P         & $3s^23p^3$    & GW  & 2$\times$ 1.9 & 2$\times$ 1.9 & 2$\times$ 2.0 & 1$\times$2.0 &2.0 \\
\end{tabular}
\end{ruledtabular}
\end{table}

\begin{table}
\caption{\label{Tab.III.2} Experimental lattice constants $a_0^{\rm exp.}$, extrapolated to $T=\SI{0}{\kelvin}$, as well as the experimental lattice constants $a_0^{\rm exp.-ZPV}$ and experimental bulk moduli $B_0^\mathrm{exp.-ZPV}$ corrected for ZPV effects. The lattice constants and bulk moduli are given in \si{\angstrom} and \si{\giga\pascal}, respectively. The Strukturbericht symbols in the parentheses indicate crystal structure: A4 = diamond, B3 = zinc-blende.}
\begin{ruledtabular}
\begin{tabular}{llll}
system    &  $a_0^{\rm exp.}$                 & $a_0^{\rm exp.-ZPV}$              & $B_0^\mathrm{exp.-ZPV}$\\
\midrule
C (A4)    & $3.567\cite{staroverov2004tests}$ & $3.553\cite{schimka2011improved}$  & $455\cite{schimka2011improved}$ \\
BN (B3)   & $3.616$                           & $3.601$                            & $410\cite{schimka2011improved}$ \\
BP (B3)   & $4.536$                           & $4.523$                            & $168\cite{schimka2011improved}$ \\
Si (A4)   & $5.430\cite{staroverov2004tests}$ & $5.421\cite{schimka2011improved}$  & $101\cite{schimka2011improved}$ \\
\end{tabular}
\end{ruledtabular}
\end{table}

The experimental lattice constants of C and Si, extrapolated to $T=\SI{0}{\kelvin}$ using the linear thermal expansion coefficients, were taken from Ref.~\onlinecite{staroverov2004tests}. For BN and BP, the experimental lattice constants measured at $T=$ \SI{300}{\kelvin} were adopted from Refs.~\onlinecite{madelung2004semiconductors}, \onlinecite{LandoltBornstein2001:sm_lbs_978-3-540-31355-7_6}, and \onlinecite{LandoltBornstein2001:sm_lbs_978-3-540-31355-7_19}, and extrapolated to zero temperature based on PBE free-energy calculations. Finally, the experimental lattice constants were corrected for zero-point vibrational (ZPV) effects. The ZPV corrections were adopted from Ref.~\onlinecite{schimka2011improved}.
We note that the BN and BP experimental lattice constants reported in Ref.~\onlinecite{haas2009calculation} were not extrapolated to $T=$ \SI{0}{\kelvin} and we note a mistake in the experimental BN lattice constant reported in Ref.~\onlinecite{haas2009calculation}, which was propagated to later works (Refs. \onlinecite{gruneis2010second}, \onlinecite{taheridehkordi2023phaseless}, and \onlinecite{schimka2011improved}).
Thus, to verify the ZPV corrections for BN and BP reported in Ref.~\onlinecite{schimka2011improved}, we re-evaluated them at the level of PBE and found identical results.
The experimental bulk moduli, corrected for ZPV effects, were taken from Ref.~\onlinecite{schimka2011improved}. Table~\ref{Tab.III.2} summarizes the crystal structures, the experimentally measured lattice constants extrapolated to $T=\SI{0}{\kelvin}$, and the ZPV-corrected lattice constants and bulk moduli.

The equilibrium lattice constants and bulk moduli were obtained by fitting the Murnaghan equation of state to the energy--volume curves, with the unit-cell volume varied in \SI{5}{\%} increments up to $\pm$\SI{15}{\%} around the experimental equilibrium volume.

\subsection{Workflow}

To accelerate the convergence of the total AFQMC energy, we hierarchically combined mean-field (HF/EXX), correlated (MP2/RPA), and stochastic (AFQMC) methods.
The key idea is that correlation energy differences between AFQMC and lower-level correlated methods converge faster than the absolute AFQMC correlation energy. Specifically, we considered MP2 using canonical HF orbitals and RPA using PBE~\cite{perdew1996generalized} orbitals as lower-level references.

The workflow can be summarized as follows: i) calculate k-point and basis-set converged MP2/RPA reference energies in the primitive cell, then ii) subtract twist-average MP2/RPA energies calculated in a supercell, and iii) add back the twist-average AFQMC energies obtained in the same supercell. This three-step scheme can be compactly expressed as follows:

\begin{equation}
    \centering
    \begin{split}
        E^{\rm X+AFQMC} =&~ E_\textrm{prim}^{\rm X}(\bm{k}\rightarrow\infty, E_{\rm cut}\rightarrow\infty) \\ &- E_\textrm{super}^{\rm X}({\rm ta}, E_{\rm cut}) \\ &+ E_\textrm{super}^{\rm AFQMC}({\rm ta}, E_{\rm cut}),
    \end{split}
    \label{Eq.III.2}
\end{equation}
where X refers to the reference correlated method --- MP2 or RPA --- "ta" denotes energies obtained via twist averaging, and "prim" and "super" refer to calculations in the primitive and supercell, respectively. The energy differences between MP2/RPA and AFQMC calculated in the supercell result in a linear shift of the equation of state of the respective reference calculation. To obtain meaningful energy differences between the supercell calculations, we ensured a consistent setup for the MP2/RPA and AFQMC --- including PW cutoffs, extrapolation with respect to $E_\mathrm{cut}^\chi$, treatment of PAW one-center terms, and FFT grids.

\subsection{Computational setup}

\begin{table}
\caption{\label{Tab.III.3}List of PW cutoffs $E_{\rm cut}$, auxiliary PW cutoffs $E_{\rm cut}^\chi$, and $\Gamma$-centered k-point grids used to sample the primitive cell for the HF, EXX, dMP2, SOX, and RPA calculations, respectively. The PW cutoffs are given in \si{\electronvolt}. $E_{\rm cut}^\chi$ refers to the largest cutoff used in the extrapolation to the infinitely-large auxiliary cutoff.}
\begin{ruledtabular}
\begin{tabular}{llll}
        & $E_{\rm cut}$ & $E_{\rm cut}^\chi$    & k-points              \\
\midrule
HF      & 650           & /                     & $12\times12\times12$  \\
EXX     & 650           & /                     & $12\times12\times12$  \\
dMP2    & 650           & 433                   & $8\times8\times8$  \\
SOX     & 650           & 433                   & $4\times4\times4$  \\
RPA     & 650           & 433                   & $8\times8\times8$   
\end{tabular}
\end{ruledtabular}
\end{table}
First, the mean-field contributions --- HF and exact exchange (EXX) --- were independently converged. Then, MP2 or RPA calculations were converged in the primitive cell with respect to both basis-set size and Brillouin-zone sampling. For RPA, VASP's built-in scheme extrapolates the correlation energy to infinitely-large auxiliary PW cutoffs $E_{\rm cut}^\chi$~\cite{harl2008cohesive}. MP2 correlation energies are manually extrapolated based on four calculations with equally spaced values of $E_{\rm cut}^\chi$; the direct (dMP2) and the second-order exchange (SOX) contributions were converged separately with respect to the k-point grid. Table~\ref{Tab.III.3} summarizes the PW cutoffs, the largest employed auxiliary PW cutoffs, and k-point grids used in the HF, EXX, MP2, and RPA calculations performed in the primitive cell.
For a more detailed discussion on how to obtain accurate RPA or MP2 correlation energies, we refer the interested reader to Refs.~\onlinecite{harl2010assessing},~\onlinecite{humer2022}, and~\onlinecite{gruneis2010second}.

Next, AFQMC correlation energies were computed in supercells containing 8, 16, and 32 atoms to progressively remove finite-size effects. We sampled the Brillouin zone by twist averaging over a generalized regular k-point mesh, equivalent to a $4\times4\times4$ Monkhorst--Pack grid in the primitive cell. This resulted in 4, 2, and 2 irreducible k-points for the 8-, 16-, and 32-atom cells, respectively. The twist-averaged (ta) energy is
\begin{equation}
    E_\text{super}^\text{AFQMC}(\text{ta}) = \sum_{\bm{k}_i} w_{\bm k_i} E_\text{super}^\text{AFQMC}(\bm{k}_i)
    \label{Eq.III.1}
\end{equation}
where $w_{\bm k_i}$ denotes the multiplicity of the irreducible k-point. Since the energy differences between the MP2/RPA and AFQMC are found to be virtually linear in the volume, it is, in principle, sufficient to evaluate them with a linear interpolation based on two or three volumes. For the 8- and 16-atom cells, we performed AFQMC calculations for all seven volumes. To reduce the computational cost for the 32-atom cells, we relied on linear interpolation based on AFQMC calculations for the smallest and largest volumes.

\begin{table}
\caption{\label{Tab.III.4}PW cutoffs $E_{\rm cut}$ and auxiliary PW cutoffs $E_{\rm cut}^\chi$ used in the supercell MP2, RPA, and AFQMC calculations. All PW cutoffs are in \si{\electronvolt}. No extrapolation with respect to $E_{\rm cut}^\chi$ was applied. The k-points column specifies the number of irreducible k-points used to sample the 8-, 16-, and 32-atom cells, respectively.}
\begin{ruledtabular}
\begin{tabular}{llll}
        & $E_{\rm cut}$ & $E_{\rm cut}^\chi$    & k-points  \\
\midrule
C       & 450           & 1012                  & 4, 2, 2     \\
BN      & 450           & 1012                  & 4, 2, 2     \\
BP      & 350           & 787                   & 4, 2, 2     \\
Si      & 250           & 562                   & 4, 2, 2
\end{tabular}
\end{ruledtabular}
\end{table}

Table~\ref{Tab.III.4} summarizes the PW cutoffs used in the supercell calculations. In AFQMC, unlike in MP2 or RPA, the correlation energy cannot be directly extrapolated with respect to $E_\textrm{cut}^\chi$, since it is recovered as a part of the Hartree and exchange contributions of the non-interacting auxiliary systems. To circumvent this, we set $E_\textrm{cut}^\chi$ equal to the kinetic energy associated with the largest reciprocal lattice vector in the FFT grid. This means that we do not employ any cutoffs for the auxiliary PW basis used to evaluate Hartree and exchange energies. A similar treatment is applied to the supercell MP2 and RPA calculations to ensure compatibility.
\begin{table}
\caption{\label{Tab.III.5} Convergence of the AFQMC equilibrium lattice constant $a_0$ (in \si{\angstrom}) and bulk modulus (in \si{\giga\pascal}) for C as a function of the PW cutoff $E_{\rm cut}$ and the auxiliary cutoff $E_{\rm cut}^\chi$. All PW cutoffs are given in \si{\electronvolt}. The AFQMC correlation energy was obtained using an 8-atom cell, along with twist averaging, and combined with the HF energy calculated in the primitive cell utilizing k-point sampling.}
\begin{ruledtabular}
\begin{tabular}{llll}
    $E_{\rm cut}$   & $E_{\rm cut}^\chi$    & $a_0^\mathrm{AFQMC}$  & $B_0^\mathrm{AFQMC}$\\
    \midrule
    450             & 1012                  & 3.5776(5)             & 441(3)\\
    550             & 1237                  & 3.5770(6)             & 441(3)\\
    650             & 1462                  & 3.5774(6)             & 441(3)
\end{tabular}
\end{ruledtabular}
\end{table}
As an illustration, Table~\ref{Tab.III.5} shows the convergence of the AFQMC equilibrium lattice constant of C with respect to the PW cutoff. In this case, no embedding [Eq.~\eqref{Eq.III.2}] was used, allowing the convergence of the AFQMC correlation energy to be isolated with respect to basis-set size. AFQMC correlation energies calculated in the 8-atom cell, along with twist averaging, were combined with basis-set converged HF energies obtained in the primitive cell utilizing k-point sampling. The resulting lattice constants and bulk moduli agree within the statistical uncertainties.

All AFQMC results presented in this work were obtained using ph-AFQMC; for simplicity, the abbreviation ``AFQMC'' refers to the phaseless variant unless stated otherwise. To reduce the computational load, the MP2- and RPA-based workflows used the same set of AFQMC calculations. For the AFQMC simulations, we employed 2560 walkers using a single-determinant HF trial wavefunction $\Psi_\mathrm{T}$. The impact of the trial density on the AFQMC total energies and the equilibrium lattice constant is discussed in Sec.~\ref{Sec.IV.B}. A comparatively large imaginary time step of $\SI{0.005}{\electronvolt}^{-1}$ was used to reduce the computational cost, as we verified that the resulting time-step error has a negligible effect on the equilibrium lattice constant (see Sec.~\ref{Sec.IV.A}). After the twist average [Eq.~\eqref{Eq.III.1}], the AFQMC total energies were converged to below $\SI{5}{\milli\electronvolt}$ per atom.
As we have demonstrated in a previous work (Ref.~\onlinecite{taheridehkordi2023phaseless}), the relative error of the AFQMC total energy scales as $\frac{1}{\sqrt{N_{\bm{k}}}}$, where $N_{\bm{k}}$ is the number of k-points. This allowed us to reduce the total number of samples by a factor $\sqrt{2}$ each time the supercell size was doubled, while maintaining a constant relative error.

Overall, the computational setup described above ensures that both mean-field and correlated quantities are converged to within \SI{0.1}{\%} for the lattice constants, providing a reliable foundation for the results presented in Sec.~\ref{Sec.IV}.

\section{Results}\label{Sec.IV}
In this section, we discuss the dependence of the AFQMC relative energies on i) the imaginary time step, and ii) the choice of exchange-correlation functional used to generate the trial wavefunction. Finally, we present the AFQMC lattice constants and bulk moduli of C, BN, BP, and Si, obtained along the workflow discussed in Sec.~\ref{Sec.III}.

\subsection{Time-step dependence}\label{Sec.IV.A}

\begin{figure}
    \centering
    \includegraphics[width=1\columnwidth]{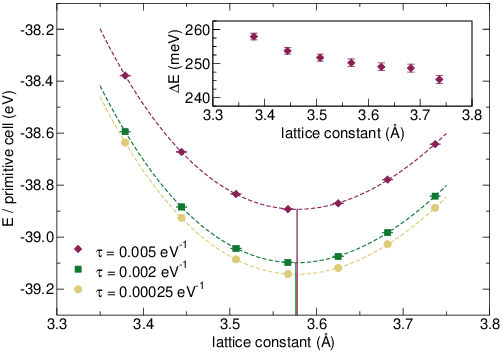}
    \caption{Equation of state of diamond calculated with AFQMC for three different imaginary time steps $\tau$. Total energies were obtained by combining twist-averaged AFQMC correlation energies (computed in an 8-atom cell and normalized to the primitive cell) with k-point converged HF energies from the primitive cell. The vertical lines connecting to the fitted curves mark the respective equilibrium lattice constants. The inset shows the total-energy differences between the largest and smallest time steps.}
    \label{Fig.IV.A.1}
\end{figure}

Figure~\ref{Fig.IV.A.1} shows the equation of state of C for different imaginary time steps $\tau =$ \SI{0.00025}{\electronvolt^{-1}}, \SI{0.002}{\electronvolt^{-1}}, and \SI{0.005}{\electronvolt^{-1}}. The total energies were calculated by adding the AFQMC correlation energies and the k-point converged HF energies. The AFQMC correlation energies were computed in an 8-atom cell using twist averaging and then normalized to the primitive cell. The HF energies were calculated directly in the primitive cell. No embedding, as discussed in Eq.~\eqref{Eq.III.2}, was employed. As expected, the total energies exhibit a systematic bias with increasing imaginary time step, amounting to about \SI{125}{\milli\electronvolt} per atom between the smallest and largest values of $\tau$. The corresponding energy differences (see inset in Figure~\ref{Fig.IV.A.1}) show a slight increase of the time-step error with increasing cell volume. This, however, has only a minor impact on the equilibrium lattice constants (\SI{3.5769(3)}{\angstrom}, \SI{3.5761(3)}{\angstrom}, and \SI{3.5776(3)}{\angstrom}, respectively), which differ by less than \SI{1.5}{\milli\angstrom}. The resulting bulk moduli (\SI{443(1)}{\giga\pascal}, \SI{441(1)}{\giga\pascal}, and \SI{442(1)}{\giga\pascal}, respectively) agree within their statistical uncertainties. 

These results suggest that structural properties are insensitive to the imaginary time step, consistent with our earlier findings for the equilibrium geometry of the N$_2$ dimer~\cite{sukurma2024toward}.
To reduce the computational cost while maintaining accuracy, we therefore chose to perform all following AFQMC calculations using an imaginary time step of $\tau =$ \SI{0.005}{\electronvolt^{-1}} and omitted the time-step extrapolation.

\subsection{Functional dependence}\label{Sec.IV.B}

\begin{figure}
\centering
    \includegraphics{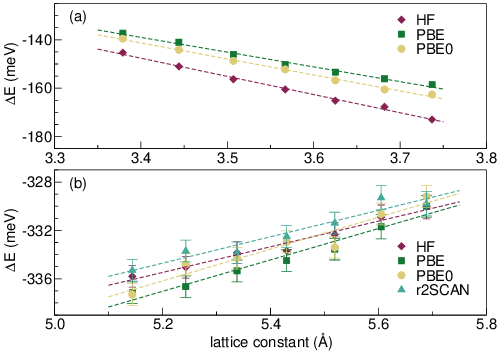}
    \caption{Energy difference per atom between AFQMC and MP2 total energies for (a) C and (b) Si, using trial wavefunctions from different exchange-correlation functionals. Calculations were performed in an 8-atom supercell using twist averaging.}
    \label{Fig.IV.B.1}
\end{figure}

We study the changes of the AFQMC equation of state using trial wavefunctions generated from the HF and PBE0~\cite{perdew1996rationale,ernzerhof1999assessment,adamo1999toward} hybrid functionals, the PBE~\cite{perdew1996generalized} generalized gradient approximation (GGA) functional, and the meta-GGA functional r2-SCAN~\cite{furness2020accurate}. We investigated the r2-SCAN functional only for Si, because the C pseudopotential lacks the required kinetic energy density.
Figure~\ref{Fig.IV.B.1} shows the energy differences per atom between AFQMC and MP2 for (a) C and (b) Si, calculated in an 8-atom supercell using twist averaging.

\begin{figure*}[t]
\centering
    \includegraphics{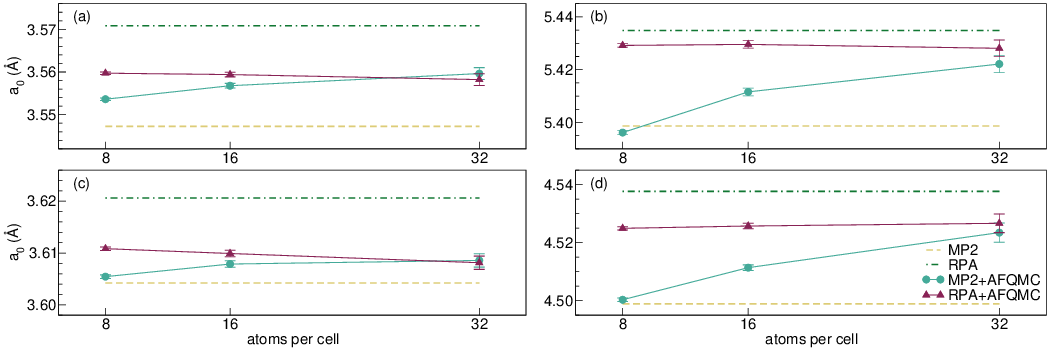}
    \caption{Convergence of the MP2+AFQMC and RPA+AFQMC lattice constants (in \si{\angstrom}) with respect to the number of atoms in the supercell for (a) C, (b) Si, (c) BN, and (d) BP. Each point corresponds to a twist-averaged AFQMC calculation embedded in MP2 or RPA calculations obtained in the primitive cell. The yellow dashed lines and the green dashed-dotted lines denote the MP2 and RPA reference lattice constants, respectively.}
    \label{Fig.IV.C.1}
\end{figure*}

For C (panel (a)), the total energies obtained with PBE and PBE0 trial wavefunctions are in close agreement, whereas the HF wavefunction yields energies lower by approximately 10\textendash15~\si{\milli\electronvolt} per atom.
This translates to only minimal changes of the equilibrium lattice constants of \SI{3.5377(3)}{\angstrom} (HF), \SI{3.5374(4)}{\angstrom} (PBE), and \SI{3.5375(3)}{\angstrom} (PBE0).
For Si (panel (b)), the investigated functionals yield total energies agreeing within 2\textendash3~\si{\milli\electronvolt} per atom.
Consequently, the equilibrium lattice constants are in close agreement \SI{5.3963(7)}{\angstrom}, \SI{5.3959(7)}{\angstrom}, \SI{5.3973(7)}{\angstrom}, and \SI{5.3957(7)}{\angstrom} using HF, PBE, PBE0, and r2-SCAN trial wavefunctions, respectively.

The observed differences in the AFQMC total energies are likely due to the bias introduced by the phaseless approximation, which guides the random walk based on a trial wavefunction. This bias will only vanish in the case of the exact many-body wavefunction.
Our findings suggest that the phaseless bias is very much system-dependent. While for Si, we observe only a mild dependence of the total AFQMC energy on the choice of trial wavefunction, for C, we notice a significant energy difference between the HF and PBE or PBE0 trial wavefunctions. 

Since AFQMC is non-variational, it is, without an accurate reference, not clear which functional yields the most accurate total energies.
Gubler \emph{et al.} recently assessed the accuracy of electronic densities for a wide range of exchange-correlation functionals~\cite{gubler2025accuracy}. According to their metrics, r2-SCAN and PBE0 produce the most accurate and reliable densities, whereas HF densities are generally less accurate~\cite{gubler2025accuracy}. This suggests that in the case of C, the HF trial wavefunction is inferior to PBE and PBE0 and thus exhibits a larger phaseless bias.

Summarizing, Figure~\ref{Fig.IV.B.1} shows that the investigated trial wavefunctions yield slightly different slopes relative to the MP2 reference.
Despite this starting point dependence, our results demonstrate that the equilibrium geometries are largely insensitive to the choice of trial density. Hence, all subsequent AFQMC calculations were performed using a single-determinant HF trial wavefunction.

\subsection{Size dependence}\label{Sec.IV.C}

Figure~\ref{Fig.IV.C.1} shows the convergence of the MP2+AFQMC and RPA+AFQMC lattice constants of (a) C, (b) Si, (c) BN, and (d) BP with respect to the number of atoms contained in the supercell. The dashed and dashed-dotted lines represent the reference MP2 and RPA lattice constants, respectively. In general, RPA+AFQMC using an 8-atom cell yields already a sizable correction to the RPA lattice constant, decreasing it on average by \SI{0.25}{\%}. Increasing the cell size from 16 to 32 atoms, the RPA+AFQMC lattice constants change on average only by \SI{0.03}{\%}.
On the other hand, the MP2+AFQMC lattice constant, calculated in an 8-atom cell, increases on average only by \SI{0.08}{\%} compared to the MP2 lattice constant. Increasing the cell size, we observe a steady increase in the MP2+AFQMC lattice constant, converging to the RPA+AFQMC lattice constant only using a 32-atom cell. The lattice constants from both workflows, obtained using 32-atom cells, agree within their statistical uncertainties.

This convergence behavior reflects the distinct physical approximations underlying RPA and MP2. The RPA, as a resummation of bubble diagrams, accurately captures electronic screening effects if the long-range DFT polarizabilities are close to the exact (many-body or experimental) polarizabilities. This follows from the fundamental fluctuation-dissipation theorem. It is generally agreed that DFT polarizabilities are good to very good~\cite{langreth_exchange}. However, due to the lack of higher-order exchange diagrams beyond Fock exchange, short-range interactions are generally less accurate. This results in an overestimation of the RPA lattice constant with respect to the experiment (see Figure~\ref{Fig.IV.D.1})~\cite{gruneis2009making}.
MP2, on the other hand, includes the second-order exchange diagram (SOX) but lacks any screening corrections to the bare bubble diagram (dMP2). Hence, MP2 tends to overestimate the polarizability in the long-wavelength limit and, as a result, underestimates the lattice constants (see Figure~\ref{Fig.IV.D.1}) --- especially in strongly polarizable systems, where electronic screening plays a prominent role~\cite{gruneis2010second}.

Our results demonstrate that AFQMC is capable of recovering the correlation effects neglected by the respective reference method and converges to a single result within statistical uncertainties, independent of the reference method. In agreement with intuition, we find that short-range correlation effects, which are missing in the RPA, are recovered in smaller cells than the long-range interactions, which are neglected in MP2. This trend is consistent across all investigated systems. Based on these findings, we identify RPA as the preferred reference method, as it results in a faster convergence of finite-size effects. The results indicate that 16-atom cells are sufficient to obtain accurate AFQMC lattice constants based on an RPA reference. Henceforth, since both workflows yield identical results within the statistical uncertainties, we report only the RPA+AFQMC results using the label ``AFQMC'' for brevity.

\subsection{Structural properties}\label{Sec.IV.D}

\begin{figure}
\centering
    \includegraphics{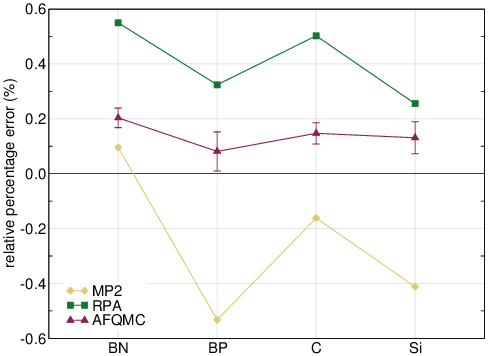}
    \caption{Relative percentage error in the MP2, RPA, AFQMC equilibrium lattice constants with respect to the ZPV-corrected experimental lattice constants at $T=$ \SI{0}{K}. MP2 and RPA results were calculated in the primitive cell utilizing k-point sampling. The AFQMC lattice constants were obtained in 32-atom cells, along with twist averaging, and using an RPA reference.}
    \label{Fig.IV.D.1}
\end{figure}

Figure~\ref{Fig.IV.D.1} shows the relative percentage error in the MP2, RPA, and AFQMC lattice constants with respect to the ZPV-corrected experimental results at $T=$ \SI{0}{\kelvin}. The reported AFQMC results are based on an RPA reference and were obtained in 32-atom cells together with twist averaging (see Sec.~\ref{Sec.III}).
On the one hand, we observe that MP2 underestimates the lattice constant by a mean absolute relative error (MARE) of \SI{0.30}{\%}. The magnitude of the relative error increases with the polarizability of the solid. This trend is most prominent along a column of the periodic table: C $\rightarrow$ Si, and BN $\rightarrow$ BP.
On the other hand, RPA overestimates the lattice constant by a MARE of \SI{0.42}{\%}. Contrary to MP2, the relative error in the RPA lattice constant decreases with the system's polarizability. The MP2 and RPA curves in Figure~\ref{Fig.IV.D.1} show parallel trends, highlighting the deficiencies of their underlying approximations.
Including AFQMC corrections consistently reduces the absolute error in the lattice constant compared to MP2 and RPA, with a MARE of \SI{0.14}{\%}. Table~\ref{Tab.IV.D.1} summarizes the calculated equilibrium lattice constants from HF, MP2, RPA, and AFQMC alongside ZPV-corrected experimental reference values for comparison.

\begin{table}
\caption{\label{Tab.IV.D.1} Calculated equilibrium lattice constants $a_0$ (in \si{\angstrom}) from HF, MP2, RPA, and AFQMC. For comparison, the experimental ZPV-corrected lattice constants, extrapolated to $T=$ \SI{0}{\kelvin}, are also listed. HF, MP2, and RPA were calculated in the primitive cell utilizing k-point sampling. AFQMC results are based on an RPA reference and were obtained in a 32-atom cell, along with twist averaging.
The associated statistical uncertainties in the last significant digit are given in parentheses. The final row reports the mean absolute relative error (MARE) with respect to the experiment.}
\begin{ruledtabular}
\begin{tabular}{llllll}
   & $a_0^\mathrm{HF}$    & $a_0^\mathrm{MP2}$    & $a_0^\mathrm{RPA}$    & $a_0^\mathrm{AFQMC}$ & $a_0^\mathrm{exp.}$   \\
\midrule
BN  & 3.596  & 3.604  & 3.621  & 3.608(1)  & 3.601 \\
BP  & 4.578  & 4.499  & 4.538  & 4.527(3)  & 4.523 \\
C   & 3.549  & 3.547  & 3.571  & 3.558(1)  & 3.553 \\
Si  & 5.498  & 5.399  & 5.435  & 5.428(3)  & 5.421 \\
\midrule
MARE (\si{\%})& 0.72  & 0.30   & 0.42      & 0.14  &
\end{tabular}
\end{ruledtabular}
\end{table}

Klime{\v{s}} \emph{et al.} showed that in the PAW method, the overlap densities between occupied and high-energy, plane-wave-like orbitals are not well described, since the PAW projectors do not accurately span the unoccupied manifold~\cite{Klimes_predictive}. However, adopting pseudopotentials with norm-conserving (NC) partial waves alleviates this issue; the RPA lattice constant decreases on average by \SI{0.3}{\%} using NC-PAWs. It is not certain whether this trend will prevail for AFQMC. Since the RPA neglects any higher-order exchange (beyond Fock exchange) and since second-order exchange will partially cancel contributions to the direct RPA, one needs to perform an explicit test for AFQMC. Using a NC-PAW for C, the resulting RPA lattice constant of \SI{3.565}{\angstrom} is \SI{0.17}{\%} smaller than the one reported in Table~\ref{Tab.IV.D.1} --- in perfect agreement with the results of Klime{\v{s}} \emph{et al.}~\cite{Klimes_singles}. However, the same NC pseudopotential yields a slight increase of \SI{0.08}{\%} in the lattice constant, which is within the statistical uncertainties of the results. Hence, it is unlikely that the used PAW potentials are the source of the residual error.

We speculate that the residual error in the AFQMC lattice constants originates from the neglect of core correlation effects since the present calculations consider only valence electrons. These effects are expected to be more pronounced in heavier systems such as BP and Si, where deeper-lying semi-core states are less tightly bound and thus more polarizable. In principle, core polarizations can be included via pseudopotentials with explicit semi-core s and p states; however, the required PW basis set must be substantially larger to accurately describe the fluctuations of the deeper-lying semi-core states, which is not computationally feasible at the moment.

High-quality reference QMC or Coupled Cluster lattice constants are scarce. However, DMC lattice constants of C~\cite{maezono2007equation} and Si~\cite{alfe2004diamond} show comparable relative errors to experiment (\SI{0.22}{\%} and \SI{0.17}{\%}, respectively).

\begin{table}
\caption{\label{Tab.IV.D.2} Calculated bulk moduli $B_0$ (in \si{\giga\pascal}) from HF, MP2, RPA, and AFQMC. For comparison, the ZPV-corrected experimental bulk moduli, extrapolated to $T=$ \SI{0}{\kelvin}, are also listed. HF, MP2, and RPA were calculated in the primitive cell utilizing k-point sampling. AFQMC results are based on an RPA reference and were obtained in a 32-atom cell, along with twist averaging. The final row reports the mean absolute relative error (MARE) with respect to the experimental results.}
\begin{ruledtabular}
\begin{tabular}{llllll}
   & $B_0^\mathrm{HF}$    & $B_0^\mathrm{MP2}$    & $B_0^\mathrm{RPA}$  & $B_0^\mathrm{AFQMC}$ & $B_0^\mathrm{exp.}$   \\
\midrule
BN  & 429    & 390    & 378    & 390(4)    & 410 \\
BP  & 175    & 178    & 168    & 171(3)    & 168 \\
C   & 496    & 456    & 434    & 448(5)    & 455 \\
Si  & 104    & 101    & 98     & 99(2)     & 101 \\
\midrule
MARE (\si{\%})& 5.2   & 2.8    & 3.9       & 2.6  &
\end{tabular}
\end{ruledtabular}
\end{table}

Table~\ref{Tab.IV.D.2} summarizes the calculated bulk moduli $B_0$ from HF, MP2, RPA, and AFQMC alongside ZPV-corrected experimental reference values~\cite{schimka2011improved} for comparison.
Compared to HF, MP2 improves the bulk moduli of C and Si but over-corrects or worsens those of BN and BP. Nevertheless, the MARE improves from \SI{5.2}{\%} for HF to \SI{2.8}{\%} for MP2. RPA, on the other hand, systematically underestimates the bulk moduli across all materials, with a MARE of \SI{3.9}{\%}. We observe that the AFQMC bulk moduli, in general, increase compared to RPA and decrease compared to MP2. While AFQMC does not systematically outperform MP2 for this property when compared to experiment, it generally follows the same qualitative trends. This is consistent with the fact that the bulk modulus is a second derivative and thus sensitive to statistical noise and residual fitting errors. We report a MARE of \SI{2.6}{\%} for the AFQMC bulk modulus.

\section{Conclusion}\label{Sec.V}
In this work, we have established a comprehensive framework for performing phaseless auxiliary-field quantum Monte Carlo (AFQMC) calculations within the projector augmented-wave (PAW) method, fully integrated into the Vienna \emph{ab initio} Simulation Package (VASP). A central technical accomplishment of this implementation is the rigorous formulation of imaginary-time propagation for non-orthogonal pseudo-orbitals, which we addressed by developing an efficient, exact inversion of the PAW overlap operator. To ensure efficient local energy evaluations, we calculate direct and exchange energies only on the plane-wave grid. Restoring the exact shape of the all-electron density corrects the neglect of PAW one-center terms. Together, these approaches enable the method to operate naturally at the basis set limit defined by the plane-wave cutoff, effectively eliminating the basis-set extrapolation errors that frequently plague wavefunction-based correlated calculations in solids. Furthermore, without the need to explicitly store the two-body Hamiltonian, the memory footprint is kept tiny, allowing the calculation to run even on slim memory nodes.

By hierarchically embedding AFQMC within MP2 and RPA workflows, we demonstrated the method’s capability to systematically correct the distinct physical deficiencies of these lower-level methods. Our results highlight that MP2 systematically underestimates lattice constants due to its failure to account for long-range electronic screening, whereas RPA tends to overestimate them by neglecting higher-order exchange diagrams beyond Fock exchange. AFQMC acts as a robust corrector for both, recovering the specific missing correlation effects required to align theory with experiment. Crucially, our finite-size scaling analysis identifies RPA as the computationally superior reference method for solid-state applications. We observed that the short-range correlations missing in RPA converge much more rapidly with respect to supercell size than the long-range screening effects absent in MP2, thereby allowing the RPA-based workflow to reach the thermodynamic limit using significantly smaller supercells.

The practical accuracy of this framework is demonstrated by the structural benchmarks obtained for C, BN, BP, and Si. The computed equilibrium lattice constants achieved a mean absolute relative error (MARE) of \SI{0.14}{\%} with respect to zero-point corrected experimental values, representing a substantial improvement over the \SI{0.30}{\%} and \SI{0.42}{\%} MAREs observed for MP2 and RPA, respectively. Furthermore, we verified the robustness of the method, demonstrating that these structural predictions are largely insensitive to the choice of the imaginary time step and the employed exchange-correlation functional used to generate the trial wavefunction. While the remaining minor residual errors are likely attributable to the neglect of core correlation effects in the current pseudopotential treatment, this work firmly establishes PAW-based AFQMC as a scalable, cubic-scaling method capable of producing benchmark-quality structural data for condensed-matter systems.

\section*{Acknowledgment}
Funding by the Austrian Science Foundation (FWF) within the project  P 33440 is gratefully acknowledged. The presented computational results have been largely obtained using the Austrian Scientific Cluster (ASC).



\appendix
\section{Basis transformations}\label{App.A}
As already discussed in Sec.~\ref{Sec.II.B.4}, the transformations from the plane-wave (PW) to the $S$-orthogonal basis and \emph{vice versa} are given by
\begin{align}
    \centering
    \mathbf{C}(\tau) &= \mathbf{C}(0) \mathbf{A}(\tau) \quad& \mathrm{orbital \rightarrow PW} \label{Eq.App.A.orbital_pw} \\
    \mathbf{A}(\tau) &= \mathbf{C}^\dagger(0) \hat{\mathbf{S}} \mathbf{C}(\tau), \quad& \mathrm{PW \rightarrow orbital} \label{Eq.App.A.pw_orbital}
\end{align}
where $\mathbf{C}(0)$ denotes the PW coefficient matrix of the $S$-orthogonal orbital basis, $\mathbf{C}(\tau)$ denotes a PW coefficient matrix, $\mathbf{A}(\tau)$ denotes an orbital coefficient matrix, and $\hat{\mathbf{S}}$ is given by Eq.~\eqref{Eq.II.B.4.S_matrix}. Inserting two complete sets of PWs, the Hamiltonian matrix elements in the orbital basis are rewritten
\begin{equation}
    \centering
    \begin{split}
        \mathbf{H}_{sr} &= \braket{\tilde \psi_s(0) \mid \hat{H} \mid \tilde \psi_r(0) } \\
        &= \sum_{\bm G \bm G^\prime} \mathbf{C}^*_{\bm{G}s}(0) \braket{\bm{G} \mid \hat{S} \hat{S}^{-1}\hat{H} \mid \bm{G}^\prime } C_{\bm{G}^\prime r}(0).
    \end{split}
\end{equation}
Thus, if the $S$-orthogonal basis is complete, one can write the propagator in the $S$-orthogonal orbital basis as
\begin{equation}
    \centering
    \begin{split}
        e^{-\tau \mathbf{H}} \mathbf{A}(\tau_0) &= e^{-\tau \mathbf{C}^\dagger(0) \hat{\mathbf{S}} \hat{\mathbf{S}}^{-1} \hat{\mathbf{H}} \mathbf{C}(0)} \mathbf{A}(\tau_0) \\
        &=\mathbf{C}^\dagger(0) \hat{\mathbf{S}} e^{-\tau \hat{\mathbf{S}}^{-1} \hat{\mathbf{H}}} \mathbf{C}(0) \mathbf{A}(\tau_0).
    \end{split}
    \label{Eq.App.A.proof}
\end{equation}
Note that matrix representations of $\hat{H}$ and $\hat{S}$ in the PW basis are denoted with an operator accent ``hat''. Importantly, if the $S$-orthogonal basis is incomplete, \emph{i.e.} $\mathbf{C}(0) \mathbf{C}^\dagger(0) \hat{\mathbf{S}} \neq \mathbb{1}$, the second equality does not hold. This can easily be seen via Taylor expansion of the first line in Eq.~\eqref{Eq.App.A.proof}.
Together, the Eqs.~\eqref{Eq.App.A.orbital_pw}, \eqref{Eq.App.A.pw_orbital}, and ~\eqref{Eq.App.A.proof} show the equivalence of propagation in the $S$-orthogonal and the PW basis
\begin{align}
    \centering
    \mathbf{A}(\tau_0 + \tau) &\leftarrow e^{-\tau \mathbf{H}} \mathbf{A}(\tau_0) \Leftrightarrow \\ 
    \mathbf{C}(\tau_0 + \tau) &\leftarrow e^{-\tau \hat{\mathbf{S}}^{-1} \hat{\mathbf{H}}} \mathbf{C}(\tau_0).
\end{align}

\nocite{*}
\bibliography{bibliography.bib}

\end{document}